 at 14.40truept
 at 17.28truept
\def\apj{ApJ}

\def\apjs{ApJS}

\def\mnras{MNRAS}

\newcount\itemno
\def \ino         { \the\itemno\global\advance\itemno by 1 }

\def \etal      {{\it et al.\ }}

\def\\{\hfil\break}

\def\spose#1{\hbox to 0pt{#1\hss}}
\def\lta{\mathrel{\spose{\lower 3pt\hbox{$\mathchar"218$}}
     \raise 2.0pt\hbox{$\mathchar"13C$}}}
\def\gta{\mathrel{\spose{\lower 3pt\hbox{$\mathchar"218$}}
     \raise 2.0pt\hbox{$\mathchar"13E$}}}

\font\eightrm=cmr8
\font\sixrm=cmr6
\def\smallup#1{\raise 1.0ex\hbox{\sixrm #1}} 
\def\smallvfootnote#1#2{\vfootnote{{\eightrm\raise 1.0ex\hbox{\sixrm#1}}}
       {\eightrm #2}} 
\documentstyle[11pt,aaspp4]{article}
\singlespace 

\begin{document}
\title{Prospects for Cosmology with Cluster Mass Profiles} 

\author{Mary M. Crone\footnote{Department of Physics \&
Astronomy, The University of Pittsburgh, Pittsburgh, PA 15260},
Fabio Governato\footnote{
Astronomy Department, University of Washington, Seattle, WA 98195},
Joachim Stadel$^2$, and Thomas Quinn$^2$} 

\begin{abstract}
We test the precision with which weak lensing data can 
provide characteristic cluster mass profiles within  Cold Dark Matter (CDM)
 scenarios.  
Using a parallel treecode to simulate 
volumes as large as 500 $h^{-1}$ Mpc with good resolution, 
we generate samples  of large clusters within a standard CDM model 
and an open CDM
model with $\Omega_o=0.3$.  
We mock high-quality lensing data by including realistic errors, selecting 
cluster samples based on velocity dispersion,  
and fitting profiles 
within a realistic range in 
radius.  We find that a sample of ten clusters  
can determine logarithmic profile slopes 
with 1-$\sigma$ errors 
of about 7\%.  Increasing the sample size to twenty brings this error
down to less than 5\%, but this is still insufficient to distinguish 
the two models. 
However,   
measures of cluster profiles obtained with weak lensing
do place strong constraints  for
general CDM--like models of structure formation, 
and we discuss the optimal strategy for obtaining 
data samples to use for this purpose.

\end{abstract}

\section{Introduction }
Weak lensing of background galaxies provides a new method
of mapping the mass distribution of galaxy clusters.  
Such data provide a particularly straightforward
means of testing cosmological models, 
because they directly reflect the distribution of mass   
rather than that 
of galaxies or gas.
Comparison to models is considerably further simplified if one uses
{\it relative} mass
profiles only, without requiring absolute mass normalization.  Unlike
absolute masses, relative densities can be obtained without mapping the
cluster to high radii or knowing the redshifts of background galaxies 
(Kaiser \& Squires 1993). 

Cluster maps from weak lensing are collecting quickly 
thanks to wide-field CCDs, notably at the 
Canada-France-Hawaii Telescope. 
Nine clusters have published data.  The outer radius covered
is typically between 500 $h^{-1}$ kpc and 1 $h^{-1}$ Mpc (Smail \etal 1994, Fahlman \etal 1994,
Squires \etal 1995, Seitz \etal 1996),  
but some data go as far as 1.5 or 2.0 $h^{-1}$ Mpc 
(Bonnet \etal 1994, Luppino \& Kaiser 1996).
Data for other clusters, some with radii out to 2.0 $h^{-1}$ Mpc, 
will be 
available soon (Squires, personal communication). 
Most of the clusters which have been mapped are very rich clusters  
at intermediate redshifts of about 0.2,
with the notable exception of the high redshift cluster MS1054 (Luppino 
\& Kaiser  1996). 
Profile fits are published for A1689 (Tyson \& Fischer),
Cl 0024+1654 (Bonnet, Mellier, \& Fort 1994), and 2218 (Squires \etal
1996).  Logarithmic slopes for these clusters fall
in the range 1.0 -- 1.4.  

The development of mass reconstruction algorithms with which to
analyze these data is progressing as well. 
The standard method is that of Kaiser \& Squires 
(1993), which uses the coherent shear of
background galaxies.  Recent work has improved upon this method
(Bartelmann 1995, Seitz \& Schneider 1996),  and also provided 
alternatives,  
some of which include amplification as well as shear 
(Broadhurst \etal 1995, Van Waerbeke \etal 1996).  
An early study of the strengths
and limitations of weak lensing analysis is that 
by Miralda-Escud\'e (1991);  several more recent papers discuss 
the effects of seeing, noise, the limits of the
weak lensing approximation, and contamination of the source field by
cluster galaxies 
(Kaiser \etal 1995; Wilson \etal 1996, Squires \etal 1996).   
Another issue is the degree of lensing by large scale structure 
(Jaroszynsky
\etal 1990, Watanabe \& Tomita 1990, Blandford \etal 1991, Bartelmann
\& Schneider 1991, Kaiser 1992).  In the context of mapping 
clusters, 
large scale structure introduces contamination in the outskirts 
of each cluster. 
Increased understanding of each of these issues
will lead to more reliable mass maps, especially in the innermost and
outermost regions of clusters.

Previous analytic and N-body studies 
predict that within hierarchical structure formation scenarios, 
cluster mass profiles depend on cosmological parameters  
(Hoffman \& Shaham 1985; Crone, Evrard, \& Richstone
1994).  
Specifically, if the initial mass density power spectrum is 
$P(k) \propto k^n$, 
higher values of the density parameter $\Omega$ produce 
flatter profiles for a fixed value of the initial power index $n$. 
The situation is less clear for CDM models, which do not
have pure power-law initial conditions;  a recent study by 
Xu (1996) concludes that standard and low-density CDM models produce 
only slightly different profiles.  

Profiles within CDM models are
especially tricky to quantify because 
they themselves do not follow power
laws, but have a significant curvature.   Navarro, Frenk, \& White (1996)   
show that when normalized to their virial radii, 
CDM clusters over a wide range in mass
do have very similar shapes, on average.  However, if the virial radius
is unknown or only poorly known, as is the case for cluster observations, 
the curvature in CDM clusters means that the profile slope depends on the   
radial fitting region. 

The purpose of this letter is to examine the precision with which 
weak lensing data can provide characteristic mass profiles within CDM
models.  We use large N-body simulations to produce a large number of clusters 
in two CDM models.  
We subsample and project each cluster to
mock weak lensing data, and fit over a limited range
in physical radius (in $h^{-1}$ Mpc).
We also assume rough knowledge of the cluster velocity dispersion. 
Intrinsic cluster-to-cluster variance, sampling error, 
and a dependence on the absolute cluster mass all limit
the ability to determine precise characteristic profiles. 

\section{Methods and Results} 

We use simulations of two cosmological models,  
standard and open CDM, in two different
volumes of space (see Table 1), for a total of four
simulations.   
Using different volumes allows us to 
make predictions for clusters over a wide range in
velocity dispersion.  It also allows us to 
compare the two cosmologies using 
samples with the same velocity dispersion,
because the open CDM (OCDM) model contains fewer massive clusters per comoving
volume than the standard CDM (SCDM) model.

The simulations were generated by 
a parallel treecode with periodic boundary conditions (Stadel and Quinn 1997). 
They use a very large number of particles, 
as permitted by having a parallel code, 
and they have a rather small spline force softening.  
This permits us to resolve not only
the central parts of each cluster,  but also substructure
and its effects on the evolution of the cluster profiles.
A large softening (or, more generally, low resolution) can
affect the shape of the cluster profiles by  diminishing the maximum
phase density actually resolved, thus prematurely
erasing substructure falling into the cluster center   
(Moore \etal 1996).  
Timesteps were constrained to $\Delta t < 0.3 {\epsilon\over v_{max}}$, where 
$v_{max}$ is the approximate maximum speed and $\epsilon$ is the softening length. 
The parameters of these simulations, including the density parameter
$\Omega_o$, Hubble parameter $H_o$,  
the rms fluctuations on a scale of $8 h^{-1} $ Mpc ($\sigma_8$),
 box length, number of particles $N$, and softening length are 
summarized in Table 1. 
\vskip 0.2in

\centerline{\bf TABLE 1}
\smallskip
\centerline { Simulation Parameters }
\bigskip
\vbox{\hbox to \hsize{\hfil\vbox{\halign {#\hfil&&\quad\hfil#\hfil\cr
\noalign{\hrule}\cr
\noalign{\smallskip}
\noalign{\hrule}\cr
\noalign{\medskip}
Model & $\Omega_o$ & $H_o$ (${km/s\over Mpc}$) & $\sigma_8$ & length ($h^{-1} Mpc$) &  $N$ & $\epsilon \ (h^{-1} kpc$) \cr
\noalign{\smallskip}
\noalign{\hrule}\cr
\noalign{\medskip}
SCDM (large)    & \hfil 1.0 & \hfil 50 & \hfil 0.7 &  500  & $360^3$  & 50 \cr
 \hfill (small) & \hfil 1.0 & \hfil 50 & \hfil 0.7 &  \ 50 & $144^3$  & 30   \cr
OCDM (large)    & \hfil 0.3 & \hfil 75 & \hfil 1.0 &  500  & $360^3$  & 63 \cr
 \hfill (small) & \hfil 0.3 & \hfil 75 & \hfil 1.0 &  \ 75 & $144^3$  & 45    \cr
\noalign{\medskip}
\noalign{\hrule}\cr
\noalign{\medskip}}}\hfil}}

\subsection{The Cluster Sample} 

We identify clusters using a standard 
friends-of-friends algorithm, and select the twenty most massive clusters
in each simulation.  Clusters selected in this way span a wide range of dynamical
states and morphologies, from relaxed and semi-isolated,
to members of cluster associations of a few members.
In each of the small volume simulations there is one cluster much larger than
the others;  we reject these clusters to produce samples 
which are more localized 
in velocity dispersion.  The ranges in dispersion 
for each of these samples are 757 -- 1098 km/s and 1827 -- 2277 km/s for SCDM,
and 419 -- 674 km/s and 820 -- 1082 km/s for OCDM.  
The dispersions of  the 
largest clusters in the large OCDM volume match those in the small SCDM volume, allowing
us to compare dispersion-limited samples with clusters that are all well-resolved. 
Dispersions for clusters with weak lensing data 
currently range from 750 km/s (Fahlman \etal 1994) to 
2000 km/s and even higher (Tyson, Valdes, \& Wenk 1990), but are typically between 
1000 km/s and 1500 km/s. 

After selecting clusters from the N-body simulations, 
we need to project a three-dimensional region
around each simulated cluster into two dimensions.   
To obtain an accurate profile we should include  
any structure in the cluster outskirts,
the accumulated lensing of 
large-scale structure, and projections of other clusters along the line of sight.  
We find 
that structure in the immediate vicinity of the cluster is accounted for 
if we select a sphere out to an overdensity of five. 
(The profile slopes within our fitting region converge to a
 constant value at an overdensity greater than this.) 
The accumulated lensing from large-scale structure
along the line of sight has been previously estimated to 
produce shear at the level of a few percent, and therefore to affect only the 
outskirts of clusters (see the references in \S 1).  
The influence of projected clusters and groups is discussed by Cen (1996),
who concludes 
that much of the substructure associated with clusters can, especially
in open models, be associated with spurious superpositions of clusters
 along the line of sight.  We directly estimate whether cluster projections
 and large scale structure affect our profiles by cutting out cylinders
 (with length equal to that of the simulation volume) around a few clusters 
 in each volume,  and projecting these into two dimensions. 
 A complete treatment of these projection effects involves propagating
 light rays through the entire distance between the source galaxies and the observer;
 here we test whether profiles are necessarily changed by 
 structure within our simulation volumes of 50 -- 500 $h^{-1}$ Mpc.  We find that
there is little effect on profile slopes within 1 $h^{-1}$ Mpc,  the outer radius
of the profile fitting region we use in this paper.
In addition, problems introduced by superpositions on cluster {\it selection}  
can be alleviated by selecting from an X-ray catalog.   
Therefore, with a few caveats, we believe that 
we are able to obtain fairly accurate profiles for large clusters out to 
$1 / h^{-1}$ Mpc.  

To obtain the two-dimensional density profiles, we identify cluster centers 
as the most bound particle in each cluster. Since  we use no information at very
 small radii (less than 0.15 $h^{-1}$ Mpc), this choice is 
effectively equivalent to the two-dimensional density maximum 
for our purposes.

\subsection{The Profile Fits} 

We find profiles for each individual cluster by averaging densities in
bins of $log R \ (h^{-1}Mpc) = 0.1$.  
If we use all the particles in each cluster, Poisson error in each bin 
is fairly small (between  2\% and 8\%).  To mock lensing data, 
we subsample each bin to produce an  
error of 20\%.  

We characterize cluster mass profiles for each sample
with a power-law fitting function
within the region 0.15 -- 1.0 $h^{-1}$ Mpc.
The outer radial cutoff is motivated by constraints of telescope time and 
the desire to avoid contributions from 
large-scale structure (as discussed above; also see Bonnet \etal 1994). 
The reliability of data at small radii is limited by 
contamination by cluster galaxies (that is, misidentification of cluster members
as background source galaxies) and 
the need to remain at densities where the weak lensing approximation can be
used.  
The inner cutoff of 0.15 $h^{-1}$ Mpc is a fairly optimistic estimate of the
smallest radius at which reliable densities could be obtained for large 
clusters (Squires \etal
1996,  Wilson \etal 1996).   

Figure 1 shows a typical cluster profile, illustrating the quality of data
we assume for our analysis.  
There is not enough information
in each profile to use more complicated functions which   
describe CDM density profiles when all the information from simulations is 
used,  such as a broken 
power-law or Hernquist
function (Xu 1996, Navarro \etal 1996). 
Values of the reduced chi-squared indicate that a power law provides a 
good fit, except in some cases where substructure is clearly present. 

We calculate characteristic profile slopes for samples 
of 10 and 20 clusters.  For each sample,  
we draw 5000 bootstrap resamplings from our set of 
simulated clusters to estimate 
the mean profile and confidence levels. 
Figure 2 summarizes the power-law fits for each sample.   
For the  large OCDM volume, we also show fits to each individual
 cluster (inset).
Note the very wide range of values for individual clusters. 
The inset also illustrates that the variance within each sample is {\it not}
caused by a velocity dispersion dependence;  within each sample, there
is no significant trend with velocity dispersion.
The larger errors in the SCDM slopes is consistent with the 
fact that clusters in high-$\Omega$ models exhibit more local, (i.e. real) 
substructure  (Crone, Evrard, \& Richstone 1996).
The intrinsic scatter in cluster profiles highlights the difficulty in drawing 
any conclusions based on small numbers, and it may severely affect measurements like
the baryon abundance in clusters (see Loewenstein and Mushotzky, 1996). 
The error bars indicate $1-\sigma$ errors for the 10-cluster samples.
Increasing the sample size to 20 decreases errors by about 30\%, very close
to the $1/\sqrt{N}$ prediction from Gaussian statistics. 

The most striking trend in Figure 2 is the dependence of slope on velocity
dispersion. There are two causes for this.  The primary one is that the profiles
are not on average, pure power laws, so that the fixed physical range
in radius used for the fit corresponds to a different region of the
cluster depending on the cluster size.  
The other is that the profiles,
even when rescaled to a characteristic radius, such as the virial radius,
are intrinsically dependent on mass.  Previous papers indicate that the
second effect is quite small (Navarro \etal 1996, Xu 1996).  
We check this by fitting the 2000 km/s clusters to a new
radial range determined by rescaling to 
approximate virial radii of the 900 km/s clusters.   
The new slope is consistent with
that for the 900 km/s clusters in both models.
Therefore, rescaling the fitting radius this way 
cannot be used to distinguish OCDM from SCDM, but the similarity of
 the two models, can be used to actually put a strong constraint on the 
whole class of CDM--like models.

If we compare  the cluster samples in the 1000 km/sec velocity range (the only one in common
between the two cosmological models  and the simplest to assemble observationally), 
we find that OCDM and SCDM cannot be distinguished at the 2-$\sigma$ level,
even using a 20-cluster sample. 

We tried unsuccessfully to improve  the cosmological signal at higher radii by fitting 
within the larger region 0.15 -- 2.0 $h^{-1}$ Mpc. 
The logarithmic slopes systematically shift steeper
by about 0.15, but the errors do not significantly decrease. 
This not entirely surprising, because this is not a huge increase in logarithmic
radius ---  the density changes by only a factor of
two, and we gain only three data points.  It may be possible to get better data
within this range (at the expense of significant telescope time);  for an even 
distribution of source galaxies and an isothermal profile, signal-to-noise is constant
as a function of radius. 
However, it may not be worth the effort to do so (for this specific
purpose) because both nearby substructure and intervening large-scale structure 
wreak havoc with the profiles at these relatively low overdensities.  Also,
it is more straightforward to compare fits which were made using constant error
bars, because otherwise one region of the cluster is 
weighted more heavily than others. 

This profile curvature
suggests that it might be appropriate to attempt fitting
to a more complicated function, if not for each individual cluster, 
for an average profile for the sample.   However, we find that for these
samples the errors in such multi-parameter fits are large enough that it
is more useful to use a simple power law.
 
\section{Discussion} 

Within the CDM models considered here, a sample of ten clusters with 
high-quality lensing data can determine logarithmic 
profile slopes with 1-$\sigma$ errors of
about 7\%.  Increasing the sample size to twenty lowers the errors to 
about $5\%$. 
This is insufficient to distinguish the models, and additional data at larger radii
 do not help. 
Nevertheless,  profiles for samples this large put a strong constraint on
CDM models as a whole, independently of the particular model adopted.
This is especially important as recent data challenge not only the SCDM model
(Loewenstein \& Mushotzky, 1996) but the whole class of CDM-like models
(Davis \etal 1992, White \etal 1993).
Moreover cluster profiles  have the major advantage of not requiring  
any absolute mass normalization  and  are  easily obtainable with present
day telescopes.

We find that it is crucial to perform fits within a consistent radial range,
and to use a cluster sample within a limited range in velocity dispersion.
This is because the profiles are not exact power laws, so that 
a fixed aperture corresponds to a different characteristic region of the cluster, 
depending on its size. 
Another point to consider is that  
clusters with lensing data are scattered over a range of redshift space,
while our simulated clusters are at $z = 0$, although significant evolution
is not expected (Xu 1996).

Although this study was designed with weak lensing data in mind,
the results can also be compared with other measures of cluster
profiles, such as X-ray maps and galaxy distributions (for example,
Lubin \& Postman 1996). We conclude that an ideal observational program should involve  
cluster selection using an X-ray catalogue (Luppino \& Gioia 1994), 
with further cuts to roughly limit the sample in both redshift and velocity dispersion. 
Such data would also form a base for systematic studies of mass-to-light ratios
and substructure,  providing powerful and direct constraints on the structure of 
clusters  (Wilson, Cole \& Frenk 1996b).

\acknowledgments 
We thank George Lake, Gordon Squires, and Gus Evrard for useful discussions. 
The simulations were performed at the ARSC, NCSA, PSC, and CTC supercomputing centers. 
This research was funded by the NASA HPCC/ESS program. 

\vfill\eject
\null

\clearpage

\begin{figure}
\epsscale{1.0}
\vskip -1.0in
\plotone{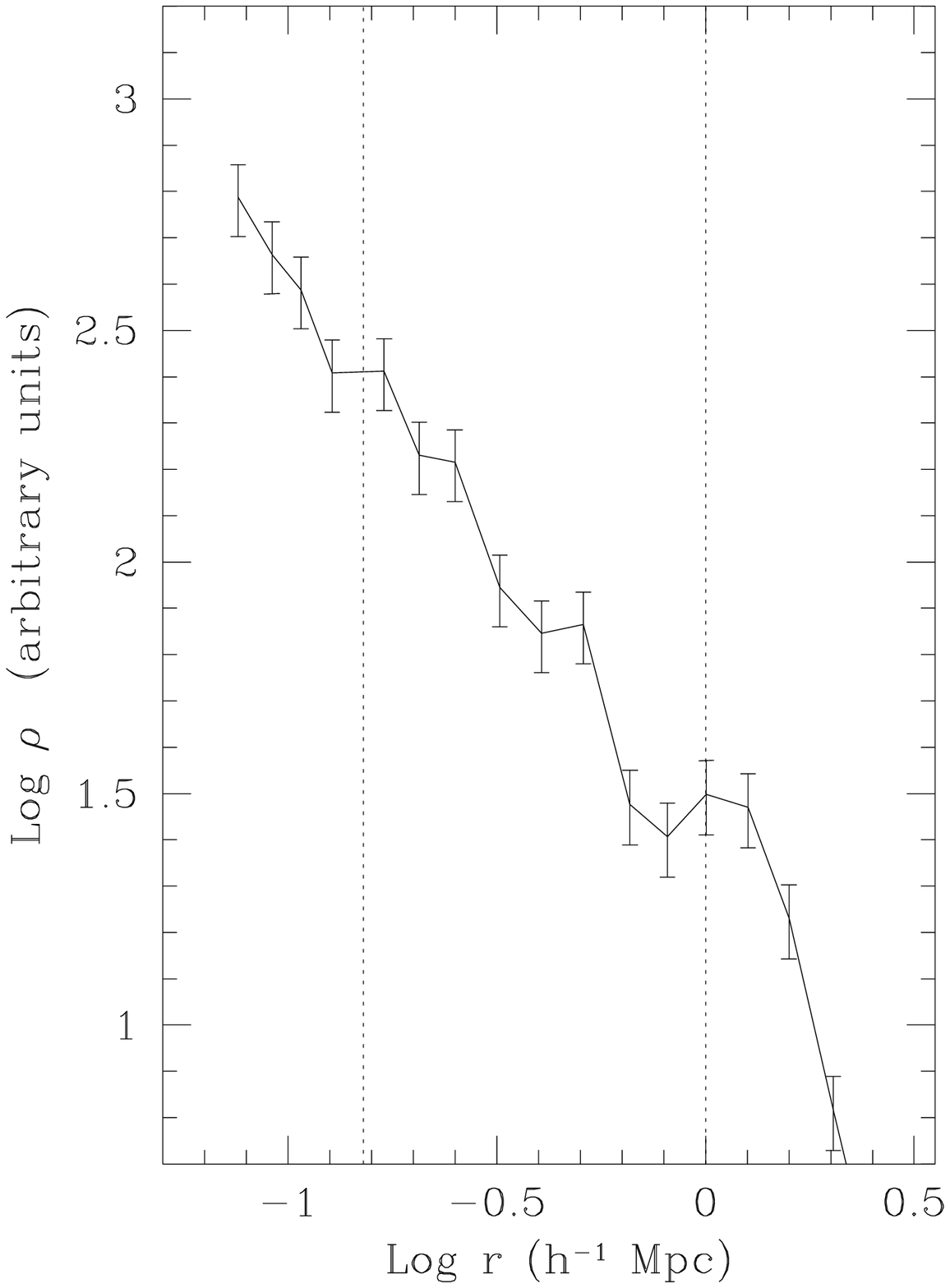}
\caption{A typical profile within the SCDM model, illustrating the quality of
data we assume for our analysis.  Dotted vertical lines indicate the
radial region within which we perform our
fits:
$0.15 -- 1.0 \ h^{-1}$ \ Mpc.
Power laws provide good fits to the cluster profiles, except in some
cases where substructure is clearly present.}
\end{figure}
	      
\begin{figure}
\epsscale{1.2}
\vskip -1.0in
\plotone{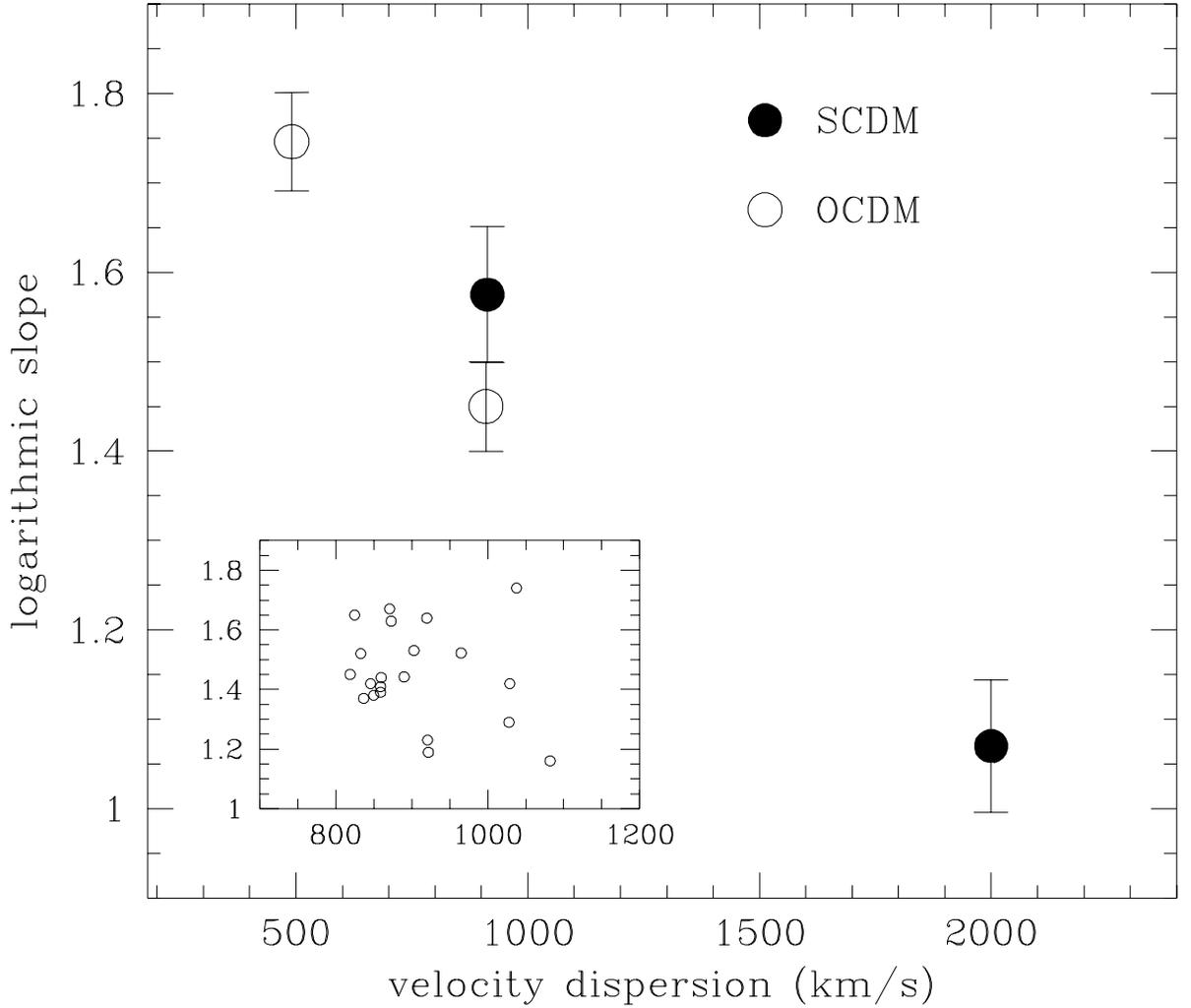}
%\vskip -1.5in
\caption{Best fit power-law slopes versus the average one-dimensional
velocity dispersion of
each sample, in the radial
range 0.15 -- 1.0 $h^{-1}$ Mpc.   We also show
slopes and velocity dispersions for each individual cluster in the sample of larger
OCDM clusters (inset).  The range of dispersions for each sample is given in the
text.
All error bars indicate $1-\sigma$
errors for the 10-cluster samples.
}
\end{figure}

\end{document}